\documentclass[%
 reprint,onecolumn,
 amsmath,amssymb,
 aps,
]{revtex4}

\usepackage{graphicx}
\usepackage{dcolumn}
\usepackage{bm}
\usepackage[table,xcdraw]{xcolor}
\usepackage{physics}
\usepackage{tikz}
\definecolor{fgreen}{RGB}{204,223,181}
\definecolor{arm}{RGB}{100,140,171}
\usetikzlibrary{arrows.meta, fit, shapes, positioning}

\begin{document}

\title{Coherence of multipartite quantum states in the black hole quantum atmosphere}

\author{Adam Z. Kaczmarek}\email{adamzenonkaczmarek@gmail.com}
\author{Dominik Szcz{\c{e}}{\'s}niak}
\author{Zygmunt B{\c a}k}
\affiliation{Institute of Physics, Faculty of Science and Technology, Jan D{\l}ugosz University in Cz{\c{e}}stochowa, 13/15 Armii Krajowej Ave., 42200 Cz{\c{e}}stochowa, Poland}
\date{\today}

\begin{abstract}
According to the recently introduced concept of quantum atmosphere, the black hole radiation is suggested to originate from the quantum excitations at the effective distance ($r$) near the event horizon ($r_H$). Here, this concept is explored from the quantum resource perspective by analysing the coherence of multipartite quantum systems located near a black hole. For the Greenberger–Horne–Zeilinger state, it is found that signatures of the atmosphere are apparent. This is to say, the coherence exhibits peak close to the event horizon and next decreases, recovering conventional behavior at $r/r_H \rightarrow\infty$. Interestingly, it is shown that as the quantum state gets more complex and the number of parties increases, the role of quantum atmosphere diminishes and the standard behaviour expected for the $N$-partite coherence quantifiers can be observed. That means, in case of complex setups the quantum atmosphere signatures may not be detectable. Hence, our findings show that care should be taken, regarding size of a system, when quantum atmosphere argument is considered.
\end{abstract}

\maketitle

\section{Introduction}

Despite decades of intellectual effort, the concept of the black hole (BH) is still not fully understood \cite{navarro2005,chen2015,raju2022}. Apart from the singularity at its core, black holes appear to violate the conservation of information, an issue known as the information paradox. This famous paradox states that while a black hole evaporates, some amount of information is lost along the way \cite{raju2022}. According to Hawking, the final state of BH radiation depends only on mass, electric charge, and angular momentum \cite{hawking1975,hawking1976,hawking1982}. Hence, when a black hole evaporates, some information appears to be lost along the way since many initial states may lead to the same final one, challenging the unitarity principle. In other words, without knowledge of the initial state, describing the full-time evolution of a black hole becomes de facto impossible \cite{raju2022}. Many routes and approaches to the resolution of the aforementioned issue rely on some kind of modification of the Hawking law, with the firewall argument as a leading example \cite{almheiri2013}. In principle, the information paradox can be addressed by redefining the source of Hawking radiation and the preservation of information encoded in that radiation \cite{giddings2006,kim2017}. Hence, it has been argued that in order to better understand Hawking radiation, it is necessary to answer one question: where exactly does black hole radiation originate?

Typically, the prevailing answer locates the origin of Hawking radiation in a proximity to the horizon ($\Delta r<<r_H$), implying a short distance scale that alters the semi-classical radiation spectra emitted by a black hole. Unfortunately, this is a popular misunderstanding and the proper picture is based on a significantly greater uncertainty about the region where Hawkings quanta is created. Recently, Giddings suggested that the Hawking radiation arises from the so-called quantum atmosphere, a region well beyond the event horizon of a black hole \cite{giddings2016}. From the estimation of the radiating black body size and comparison between the Hawking and the Stefan-Boltzmann law for photons, he concluded that the effective emitting area of the Hawking radiation is larger than the size defined by the radius of a black hole alone. Thus, the source of the Hawking quanta is modified by the effective event horizon with the radius $r_A$, satisfying relation $\Delta r = r_A-r_H \sim r_H$ \cite{giddings2016}. This argument was further backed up by studies of the stress-energy tensor and the Schwinger effect in the context of quantum atmosphere \cite{dey2017,dey2019}, as well as by the analysis of the charged back holes \cite{ong2020}. Moreover, the analogue of quantum atmosphere region has been found in the acoustic black holes by the analysis of quantum correlations \cite{balbinot2022}. 

Unfortunately, the potential role of quantum atmosphere in shaping quantum correlations near a black hole remains still largely unexplored. Given the intrinsic link between information and geometry, this constitutes an important and intriguing research direction that not only creates yet another opportunity for validation of the quantum atmosphere concept but also allows to explore quantum phenomena in a new relativistic setting. This is particularly important since most of the available studies considers conventional picture of the Hawking radiation that may hinder some of the local effects \cite{hu2018}. For example, recently conducted analysis shows that the nonlocal correlations exhibit strong distance-dependent signatures of the quantum atmosphere \cite{kaczmarek2023b}. Still, we argue here that even this approach may be somewhat limited due to the assumption of a bipartite states. As quantum tasks become more complex, more entangled particles are needed for their execution and one can not practically rely on the bipartite systems \cite{wu2022,wu2022B}. Additionally, multipartite states exhibit enhanced resilience against noise and related effects, thereby offering greater survivability \cite{XIAO2022}. Therefore, ensuring the longevity of a system becomes a practical necessity if one wants to test Giddings argument in a more practical manner. In this context, the quantum measure of coherence appears to be perfectly suited for addressing multipartite states experiencing Hawking radiation \cite{Wu2019}. This is due to the fact that coherence is not only recognized as a fundamental property of quantum systems but also a crucial resource for quantum computation and information processing that strongly relies on a degree of correlation between constituting states. This understanding, built upon the seminal work of Baumgratz, Cramer, and Plenio \cite{baumgratz2014}, emphasises significance of coherence in driving advancements within the field. In particular, the core of this approach was to develop measure for coherence which will be analogous to the previously established entanglement quantifiers, defining a maximally coherent state as a unit of coherence (similarly to maximally entangled states being units of entanglement). As a result, two viable distance-based quantifiers of coherence: the so-called $l_1$-norm and the relative entropy of coherence (REC) were introduced. For more detailed information about quantum coherence as a resource and its applications, please see \cite{streltsov2017}.

In the context of the above, herein we will attempt to explore the behavior of multipartite coherence under influence of Hawking quanta when taking into account Giddings argument \cite{giddings2016}. Hence, we will aim to extend previous studies that assumes bipartite states \cite{alsing2006,martinmartinez2010,lanzagorta2014,huang2017,hu2018,Wu2019,wu2020}, particularly our recent analysis that takes into account the quantum atmosphere influence \cite{kaczmarek2023b}. In details, in section II, after general description of fermionic fields in the background of a black hole we will introduce coherence measures. Next, in section III, the Greenberger–Horne–Zeilinger (GHZ) tripartite state will be analysed for both $l_1$-norm of coherence and REC in the physically accessible and inaccessible scenarios. Similar investigations will be conduced for the general multipartite coherence. Our manuscripts will end with summary and some pertinent conclusions.

\section{Dirac fields in the Schwarzschild space-time}
\label{methods1}
To study the behavior of multipartite coherence near the quantum atmosphere region, we begin with the fermionic type of initial state. In this regard, it is instructive to note that the fermionic modes are more resistant towards the Hawking radiation than the bosonic ones and have non-zero value even for the radiation temperature approaching infinity \cite{he2016, hu2018, kaczmarek2023}.

In the context of the above, the Dirac equation for the curved space-time is first considered:
\begin{align}
    (i\gamma^a e^\mu_aD_\mu-m)\psi=0,
    \label{eq1}
\end{align}
where $D_\mu=\partial_\mu-\frac{i}{4}\omega^{ab}_\mu \sigma_{a b}$, $\sigma_{ab}=\frac{i}{2}\{\gamma_a,\gamma_b\}$, $e^\mu_a$ is {\it vierbein} and $\omega^{ab}_\mu$ denotes the spin connection. In order to obtain a complete basis for the analytic modes with the positive energy, the Kruskal coordinates are utilized to perform analytical continuation in accordance to the Damour-Ruffini method \cite{damour1976, dey2017}. The resulting Dirac fields are expanded in the appropriate Kruskal basis, as follows:
\begin{align}
    \Psi=&\sum_i d\textbf{k}\frac{1}{\sqrt{2\cosh(\pi \omega_i/\kappa)}}\Big[ c^I_\textbf{k}\zeta^{I+}_\textbf{k}+c^{II}_\textbf{k}\zeta^{II+}_\textbf{k} +d^{I\dagger}_\textbf{k}\zeta^{I_-}_\textbf{k}+d^{II\dagger}_\textbf{k}\zeta^{II_-}_\textbf{k}\Big],
    \label{eq2}
\end{align}
where $c_\textbf{k}$ and $d_\textbf{k}^\dagger$ are subsequently the creation and annihilation operators applied to the Kruskal vacuum for regions $I$ and $II$ \cite{pan2008}. Next, by using the Bogoliubov transformation, it is possible to establish the relation between operators in a black hole and the Kruskal space-times \cite{ge2008}. In particular, the vacuum and excited states of the black hole coordinates correspond to the Kruskal two-mode squeezed states as:
\begin{align}  
    \nonumber &\ket{0_{\textbf{k}}}^+=\alpha\ket{0_\textbf{k}}_I^+\ket{0_{-\textbf{k}}}^-_{II}+\beta \ket{1_\textbf{k}}_I^+\ket{1_{-\textbf{k}}}^-_{II},\\&\ket{1_{\textbf{k}}}^+=\ket{1}^+_{I}\ket{0_{-\textbf{k}}}_{II}^-,
    \label{eq3}
\end{align}
with the following Bogoliubov coefficients \cite{lanzagorta2014}:
\begin{align}
    \alpha=\frac{1}{(e^{-\omega_i /T}+1)^{1/2}},\;\;\; \beta=\frac{1}{(e^{\omega_i /T}+1)^{1/2}},
     \label{eq4}
\end{align}
where $T$ denotes Hawking temperature of the emitted radiation. The $\ket{n}_I$ and $\ket{n}_{II}$ are associated with the orthogonal bases for the regions inside and outside of a black hole \cite{navarro2005,alsing2006}.

The above reasoning can be directly linked to the Giddings argument \cite{giddings2016}, by considering a dimensionally reduced Schwarzschild black hole of line element:
\begin{align}
ds^2=-f(r)dt^2 + f(r)^{-1}dr^2,
\end{align}
and substituting it into the reduced Dirac equation from Eq.(\ref{eq1}).

\begin{figure*}
    \centering
    \includegraphics[scale=0.85]{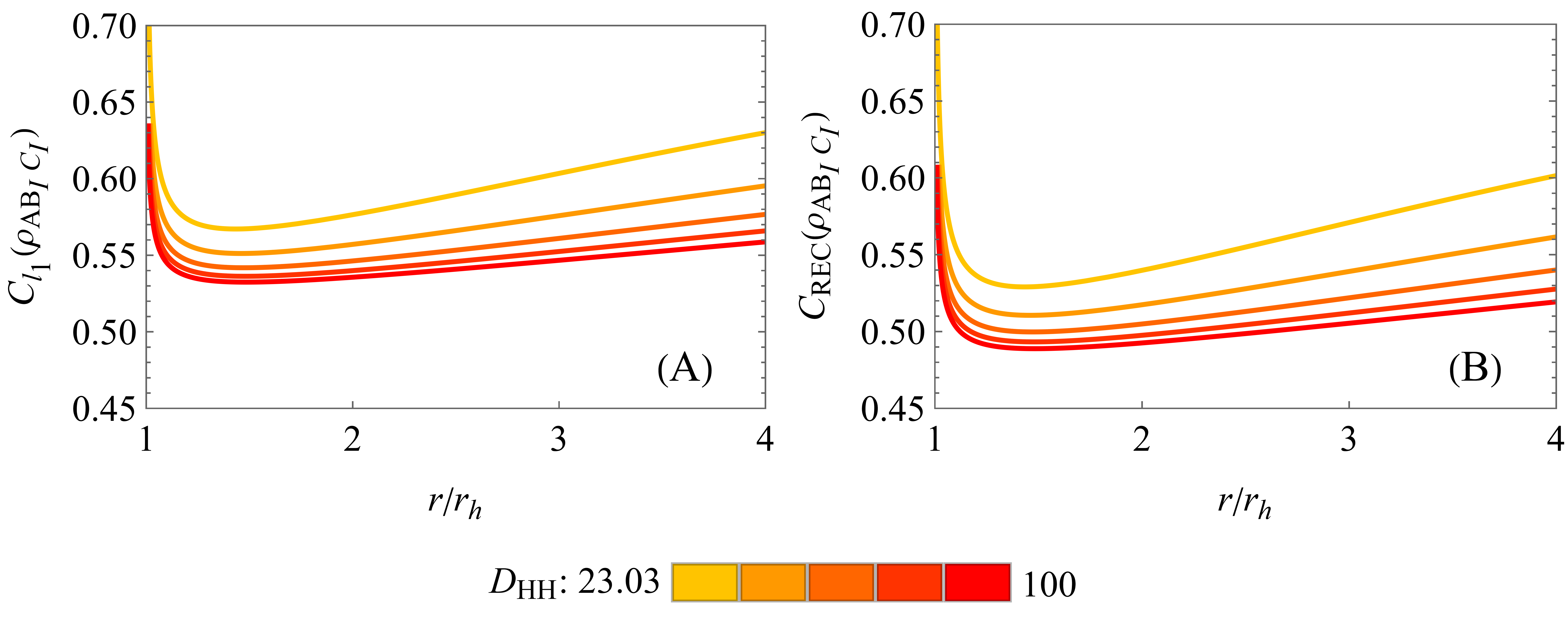}
    \caption{The physically accessible (A) $l_1$-norm of coherence and (B) relative entropy of coherence as a function of the normalized distance ($r/r_{H}$) for the selected values of the constant ($D_{HH}$) of tripartite subsytem $\rho_{AB_IC_I}$.}
    \label{r1}
\end{figure*}

Next, in order to study $N$-partite Dirac system and its coherence, we start with the N-mode extension of the GHZ state \cite{wu2022}:
\begin{align}
    \Phi_{123...N}=\big( \eta \ket{0}^{\otimes N} +\sqrt{1-\eta^2}\ket{1}^{\otimes N}\      \big)_{123...N},
    \label{eq5}
\end{align}
where the $i$-th ($i=1,2,...,N$) mode is detected by apparatus of the $O_i$-th observer, and where parameter $\eta \in [0,1]$. Furthermore, in order to study how such $N$-partite system behaves it is assumed that $n(n<N)$ observers will hover near the black hole with the uniform acceleration, in the distance $r$ from the BH centre that is extended beyond the radius $r_h$ ($r>r_h$) of an event horizon. The rest of the $N-n$ observers will remain stationary at the asymptotically flat region of spacetime. It is worth to mention that such GHZ state is not the only possible way to consider tripartite case, as multipartite W-state can also be studied but due to its complexity it is rarely done \cite{wu2020}.

In principle, according to Giddings quantum atmosphere argument \cite{giddings2016}, due to the radiation originating from that region, the modes perceived by observers near the event horizon will change to the Kruskal ones by using the Bogoliubov transformation between modes, according to the Eqs. (\ref{eq3}) and (\ref{eq4}) \cite{ge2008, pan2008}. Moreover, we want to clarify that introduced setup is a {\it{gedanken}} one and the analysis presented here is complementary to the previous works devoted to quantum correlations and quantum atmosphere argument \cite{balbinot2022,kaczmarek2023b}. In this way the density matrix, describing state for the $N$-partite system of interest, can be obtained \cite{Wu2019}.

Given the d-dimensional state in the basis $\{\ket{i}\}_{i=1,...,d}$, the associated $l_1$-norm of coherence is defined as \cite{baumgratz2014}:
\begin{align}
C_{l_1}(\rho)= \sum_{i \neq j}|\rho_{i,j}|,
\end{align}
that is, as the sum of the absolute values of all off-diagonal elements of the density matrix $\rho$. 

On the other hand, another quantifier, the relative entropy of coherence (REC), is found to be:
\begin{align}
   C_{REC}(\rho)=S(\rho_{diag})-S(\rho),
\end{align}
where $S$ denotes the von Neumann entropy and $\rho_{diag}$ is the state constructed by deleting off-diagonal elements of matrix $\rho$ \cite{baumgratz2014}.

We note, that since our goal is to characterise the influence of the quantum atmosphere on a multipartite coherence, the local temperature will take the Hartle-Hawking ($T_{HH}$) form \cite{eune2019,kaczmarek2023b}:
\begin{align}
    T_{HH}=T_H \sqrt{1-\frac{r_h}{r}}\sqrt{1+2\frac{r_h}{r}+\big(\frac{r_h}{r}\big)^2\Big(9+D_{HH}+36 \ln(\frac{r_h}{r})\Big)},
    \label{eq8}
\end{align}
with $T_H=\frac{1}{4\pi r_h}$. In Eq. (\ref{eq8}), the $D_{HH}$ is the undetermined constant of the stress tensor for the Hartle-Hawking vacuum, termed here as the Hartle-Hawking constant for simplicity. Note that this temperature expression is in accordance with the main motives of Giddings work \cite{giddings2016,eune2019}.

\section{Results}
\label{results}

\subsection{GHZ state}

For the tripartite system, we assume three distinguished observers, namely archetypical Alice ($A$), Bob ($B$) and Charlie ($C$), equipped with the suitable particle (fermionic) detectors. By convention, when referring to the observers, their names will be used, while Latin letters will be associated with their devices. In this case, Eq.(\ref{eq5}) reduces to:
\begin{align}
    \Psi_{ABC}=\eta\ket{0}_A \ket{0}_B\ket{0}_C+\sqrt{1-\eta^2} \ket{1}_A \ket{1}_B\ket{1}_C.
\label{eq9}
\end{align}
where $\ket{0}_a$, $\ket{1}_a$ with $a=A,B,C$ denote vacuum and excited states for Alice, Bob and Charlie. In this scenario, only Alice will remain in the asymptotically Minkowski region, while both Bob and Charlie will be located at a distance $r$ close to the event horizon. This can be achieved by Bob and Charlie hovering at some distance $r$ in the proximity of the BH horizon, after free fall into the direction of a BH. Thus, devices $B$ and $C$ are influenced by a radiation that can originate from the quantum atmosphere. Since of our particular interest is coherence in the physically accessible region (meaning outside the event horizon), the trace over regions $II$ for both $B$ and $C$ detectors have to be implemented. In what follows, $l_1$-norm of the physically accessible coherence for the state with two observers near the event horizon is \cite{Wu2019}:
\begin{align}
    C_{l_1}(\rho_{AB_I C_{I}})=2\eta \sqrt{1-\eta^2}\alpha^2.
    \label{eq10}
\end{align}
Similarly, the system relative entropy of coherence (REC) is found to be:
\begin{align}
C_{REC}(\rho_{AB_IC_I})=-\frac{\alpha^4 \eta ^2 \log \left(\alpha^4 \eta ^2\right)}{\log (2)}+\frac{\left(\alpha^4 \eta ^2-\eta ^2+1\right) \log \left(\alpha^4 \eta ^2-\eta ^2+1\right)}{\log (2)}-\frac{\left(1-\eta ^2\right) \log \left(1-\eta ^2\right)}{\log (2)}.
  \label{eq11}
\end{align}
Plots of $l_1$-norm of coherence and REC are presented in Fig.(\ref{r1}). Interestingly, both measures of coherence behave vastly different from the case when the quantum atmosphere argument is not invoked \cite{Wu2019,wu2020,wu2022B}. Moreover, after initial and sudden vertical drop of coherence, the values of both $l_1$-norm and REC are increasing, contrary to the results obtained for Hawking radiation that originates from the event horizon. It is argued that observed unusual characteristics of coherence may be associated with the different origin of Hawking quanta. This is evident from Fig.(\ref{r1}) since sudden loss of coherence can be noticed close to the horizon, where $r\sim r_h$. 

Another observation that differs from the previous results for bipartite systems, particularly those related to the measurement-induced nonlocality (MIN) measure, is the behavior of the $D_{HH}$ constant. Specifically, for $D_{HH}=\{23.03,40,60,80\}$ the minimal values of $C_{l_1}$ are found for the scaled distances $r/r_H \sim \{1.43,1.46,1.47,1.48\}$ accordingly. That means, the distance at which the maximum correlations value is observed is clearly shifted when the $D_{HH}$ constant changes. This is in contradiction to the results obtained previously for the MIN measure, where the position of the counterpart correlations minimum was constant \cite{kaczmarek2023b}. Notice that $D_{HH}$ present in the local temperature expression obtained by \cite{eune2019} is an integration constant, thus a mathematical {\it remnant}. Quantum coherence, as more sensible to the values given by $D_{HH}$ may be a way to constrain it from above, since lower limit of $D_{HH}\geq 23.03$ have been already introduced. However, the physical nature of this parameter remains unclear nonetheless results given in \cite{eune2019}, and its importance will diminish at the large scales. We also indicated that when moving away from the black hole horizon location  the coherence eventually {\it freezes}, since general limits of $l_1$-norm and relative entropy of coherence are:
\begin{align}
 \lim_{r/r_H \rightarrow \infty}C_{l_1}(\rho_{AB_IC_I}) = \frac{2 \eta  \sqrt{1-\eta ^2} e^{\omega }}{e^{\omega }+1}, \;\;\;\;\lim_{r/r_H \rightarrow \infty}C_{REC}(\rho_{AB_IC_I}) =\eta ^2 \log \left(\frac{\eta ^2 e^{2 \omega }}{\left(e^{\omega }+1\right)^2}\right).
   \label{eq12}
\end{align}
accordingly. Thus, in principle at large distances the conventional characteristic of coherence measures in the black hole spacetime is recovered \cite{Wu2019,wu2020}.

\begin{figure*}
    \centering
    \includegraphics[scale=0.85]{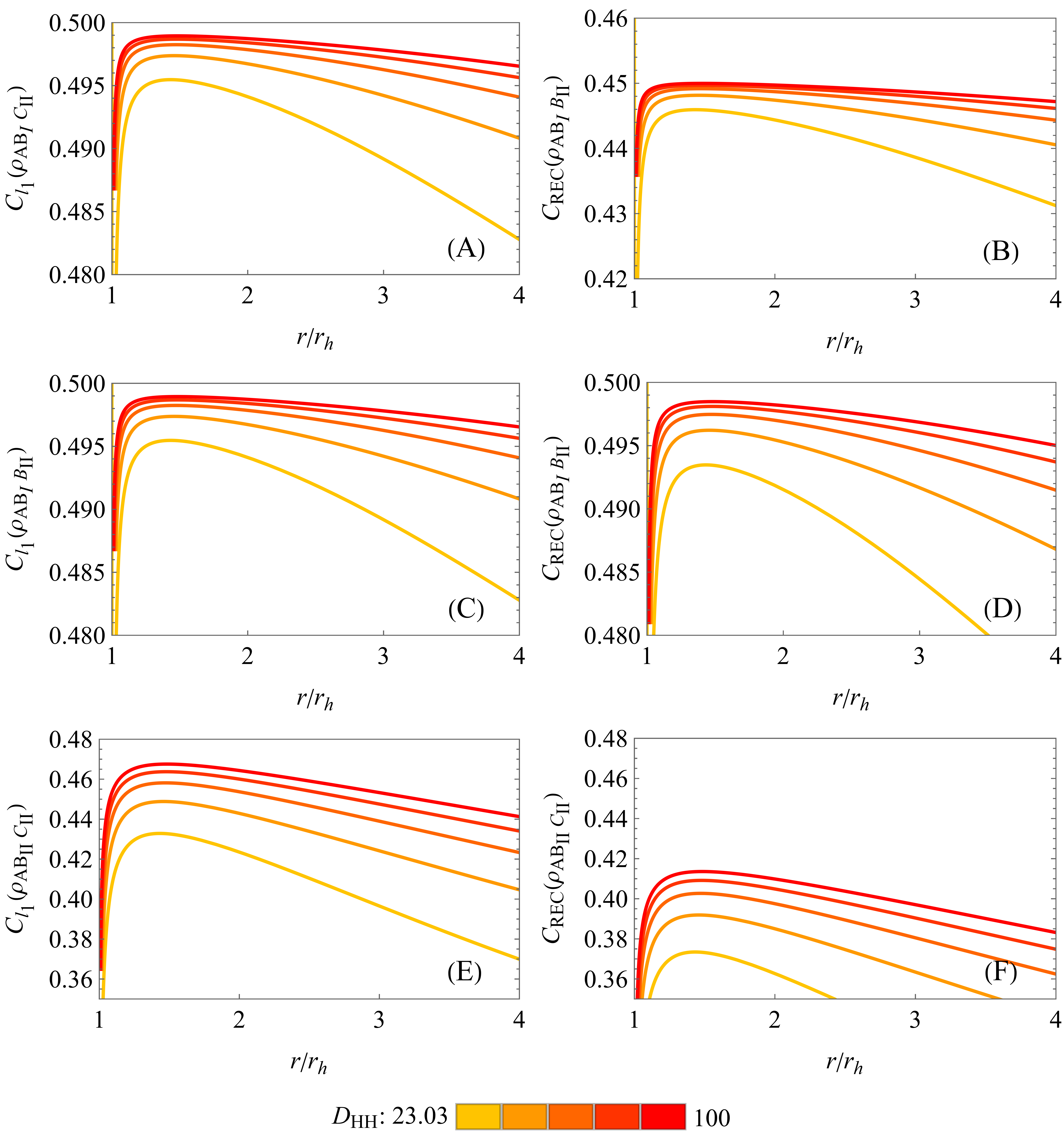}
    \caption{The $l_1$-norm of coherence and relative entropy (REC) of coherence for the physically inaccessible subsystems, that is $\rho_{AB_{I}C_{II}}$, $\rho_{AB_{I}B_{II}}$ and $\rho_{AB_{II}C_{II}}$, for different values of constants $D_{HH}$. The plots have been obtained for $\eta=1/\sqrt{2}$ and $\omega=1$.}
    \label{r2}
\end{figure*}

Finally, our attention can be turned to behavior of inaccessible subsystems, the topic hitherto unexplored in the context of the Giddings argument. The $l_1$-norm of coherence of our tripartite setup of Alice, Bob and Charlie can be also divided onto the following subsystems:
\begin{align}
  C_{l_1}(\rho_{AB_{II}C_{I}})=    C_{l_1}(\rho_{AB_IC_{II}})=2\eta \sqrt{1-\eta^2}\alpha\beta,\\  C_{l_1}(\rho_{AB_{I}B_{II}})= C_{l_1}(\rho_{AC_{I}C_{II}})=2\eta^2\alpha\beta
  \\
   C_{l_1}(\rho_{AB_{II}C_{II}})= 2\eta \sqrt{1-\eta^2}\beta^2,
\end{align}
that are nonphysical, since modes corresponding to mode $II$ locked behind an event horizon are physically inaccessible for the observers outside a black hole \cite{navarro2005,lanzagorta2014}. In a similar manner, the relative entropy of coherence in physically inaccessible scenarios reads:
\begin{align}\nonumber
C_{REC}(\rho_{AB_{II}C_I})=C_{REC}(\rho_{AB_IC_{II}})&=\frac{\left(\alpha ^4 \left(-\eta
   ^2\right)+\alpha ^2 \eta ^2-\eta ^2+1\right) \log \left(\alpha ^4
   \left(-\eta ^2\right)+\alpha ^2 \eta ^2-\eta ^2+1\right)}{\log
   (2)}\\ &-\frac{\alpha ^2 \beta ^2 \eta ^2 \log \left(\alpha ^2 \beta ^2 \eta
   ^2\right)}{\log (2)}-\frac{\left(1-\eta ^2\right) \log \left(1-\eta ^2\right)}{\log
   (2)}\\
   C_{REC}(\rho_{AB_IB_{II}})= C_{REC}(\rho_{AC_IC_{II}})&=-\frac{\alpha ^2 \eta ^2 \log \left(\alpha ^2 \eta ^2\right)}{\log
   (2)}-\frac{\beta ^2 \eta ^2 \log \left(\beta ^2 \eta
   ^2\right)}{\log (2)}+\frac{\eta ^2 \log \left(\eta ^2\right)}{\log
   (2)}\\ \nonumber
   C_{REC}(\rho_{AB_{II}C_{II}})&=\frac{\left(\alpha ^4 \eta ^2-2 \alpha ^2 \eta ^2+1\right) \log
   \left(\alpha ^4 \eta ^2-2 \alpha ^2 \eta ^2+1\right)}{\log
   (2)}-\frac{\beta ^4 \eta ^2 \log \left(\beta ^4 \eta
   ^2\right)}{\log (2)}\\&-\frac{\left(1-\eta ^2\right) \log \left(1-\eta
   ^2\right)}{\log (2)}.
\end{align}
This part of the analysis is presented in the Fig. (\ref{r2}), where $l_1$-norm of coherence and relative entropy are plotted for each inaccessible region, accordingly. Note, that general behaviour of coherence as a resource is directly opposite to the  physically accessible scenario. After a sudden increase in values, the coherence reach its maximum before decreasing monotonically. As a result the freezing effect is observed for both the $l_1$-norm and REC. Notice that smaller value of  parameter $D_{HH}$ gets leads to the stronger inaccessible solutions of coherence. Interestingly, this is opposite to the nature observed for the physically accessible scenario in case of the $AB_IC_I$ subsystem. Moreover, when physically accessible coherence increases upon shifting towards bigger radii $r/r_H$, the physically inaccessible ones are diminishing, as it can be seen in Fig.(\ref{r2}). Thus, the general behaviour of coherence as a resource for physically inaccessible subsystems under local temperature is complementary to the physically accessible ones.

\subsection{$N$-partite system}

\begin{figure*}
    \centering
    \includegraphics[scale=1]{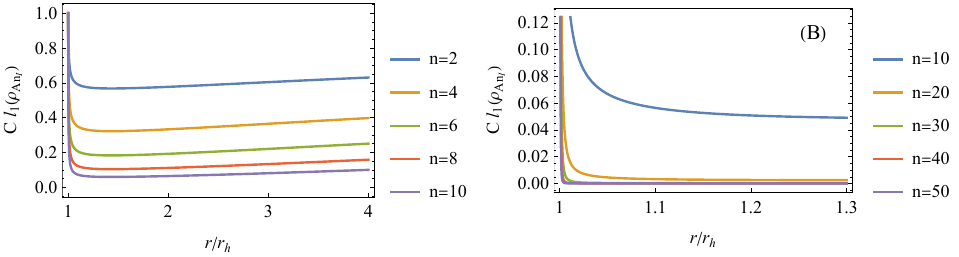}
    \caption{The behavior of the $l_1$-norm of coherence for physically accessible subsystems $\rho_{An_I}$ for different number of modes as a function of the scaled radius $r/r_h$. The figures are plotted for the $D_{HH}=23.03$, $\omega=1$ and $\eta=1/\sqrt{2}$.}
    \label{r3}
\end{figure*}

Now, it is time to consider larger system with more detectors involved. For the $N$-partite state given by Eq.(\ref{eq5}), the $l_1$-norm of coherence can take various forms that depend on the setup \cite{Wu2019}. In the present work we are interested in the scenarios with one party ({\it i.e.} $A$) that remains stationary at the flat spacetime, and $n$ observers that after a free fall begin to hover at distance $r$ near the event horizon of a black hole. It is motivated by the prospect of the presented work, where possible influence of the so-called quantum atmosphere on $N$-partite subsystems and their coherence are studied. Thus, the physically accessible $l_1$-norm of coherence takes form \cite{Wu2019}:
\begin{align}
        C_{l_1}(\rho_{N-n,n_I})=2\eta \sqrt{1-\eta^2}\alpha^n,
\end{align}
for state built from the $N-n$ Dirac fields in asymptotically flat spacetime and $n$ modes in the region near the event horizon, in accordance with (\ref{eq5}). For our considerations $N-n=1$ and corresponds to Alice that stays stationary in flat spacetime. Clearly, this relation can be considered as an extension of Eq. (\ref{eq10}). Moreover, it is easy to notice that asymptotic value of $l_1$-norm of coherence will vanish in the limit $n\rightarrow \infty$. Behavior of $C_{l_1}$ has been presented in Fig.(\ref{r3}) for the constrained value of $D_{HH}=23.03$. The general dependence of $l_1$-norm on the scaled radius $r/r_H$ remains the same, again after initial decrease the freezing phenomenon occurs for a large radii when $n$ is finite. Upon increase in the number of modes near the black hole, the physically accessible coherence decreases as a result of the power-law growth of Hawking radiation involvement. This result should be expected when invoking analysis for fermions and Schwarzshild black hole \cite{Wu2019,wu2020}.  Moreover, under increase in $n$, the signatures of quantum atmosphere are becoming less visible and evident. For the $51-$partite state ($n=50$) almost all coherence is present near the event horizon $r/r_H \sim r_H$, effectively smearing quantum atmosphere signatures visible in coherence for smaller $n$ values. This is to say, the conventional behaviour of $N$-partite coherence is recovered \cite{Wu2019,wu2020}. 


\section{Discussion and conclusions}
\label{summary}

In this work, the quantum atmosphere argument has been considered and the interplay between distinct origin of Hawking quanta with quantum coherence was studied, primarly as an extension of our previous related work \cite{kaczmarek2023}. Specifically, the system of $N$-parties have been studied together with its coherence as a resource. The $l_1$-norm of coherence and relative entropy of coherence have been analysed for the tripartite GHZ state for subsystems that are physically accessible and inaccessible, together with the physically accessible $l_1$-norm of coherence for general $N>3$-partite states as a subject to change in scaled radii $r/r_H$ and hence the scaled local temperature $T_{HH}/T_H$ \cite{eune2019,kaczmarek2023b}. For the tripartite GHZ state, the coherence behavior supports our previous findings for the MIN measure \cite{kaczmarek2023b}. The sudden change of $l_1$-norm and REC of coherence has its minimum for the $r$ well outside the event horizon $r_H$. For physically accessible (inaccessible) coherence, after such minimum (maximum) both quantifiers of coherence increase (decrease) in a monotonic way, recovering their large-distance behaviour when $r/r_H \rightarrow \infty$ seen in previous studies \cite{Wu2019}. One note should be added on monogamy of coherence for the GHZ state. Monogamy of coherence is independent on properties of the system, such as local temperature $T_{HH}$ and related  quantities. This result applies straightforwardly from previous work of Wu {\it et al.} \cite{Wu2019}.

What if the number of involved detectors is greater? Our analysis suggests that more complex system leads to the weaker influence of the black hole atmosphere. The behavior of the $l_1$-norm of coherence for $\rho_{An_I}$ subsystem changes visibly (see {\it e.g.} Fig(\ref{r3})). The coherence measure sharpens very close to the event horizon, mitigating non-trivial behavior of radiation in the quantum atmosphere region. Thus, for quantum systems and devices composed of larger number of modes the impact of the different Hawking quanta origin may be negligible, as speculated by \cite{giddings2016,eune2019,ong2020,kaczmarek2023b}. In summary, we conclude that when studying quantum correlation or coherence from the RQI point of view, one should be careful with the complexity of system since quantum coherence as a resource is notably sensitive to the increasing partition of initial state.

\bibliographystyle{apsrev}
\bibliography{bibliography}

\end{document}